\documentstyle[prl,eqsecnum,aps]{revtex}

\begin{document}

\twocolumn[

\preprint{ILC/qcomp-28Mar95/quant-ph/9505011}
\draft

\title{A Simple Quantum Computer}
\author{Isaac L. Chuang \/{\em and}\/ Yoshihisa Yamamoto}
\address{\vspace*{0.5ex}{\em ERATO Quantum Fluctuation Project} \\
	Edward L. Ginzton Laboratory, Stanford University,
		Stanford, CA 94305 }
\date{\today}
\maketitle

]


\begin{abstract}
We propose an implementation of a quantum computer to solve Deutsch's
problem, which requires exponential time on a classical computer but
only linear time with quantum parallelism.  By using a dual-rail qubit
representation as a simple form of error correction, our machine can
tolerate some amount of decoherence and still give the correct result
with high probability.  The design which we employ also demonstrates a
signature for quantum parallelism which unambiguously deliniates the
desired quantum behavior from the merely classical.  The experimental
demonstration of our proposal using quantum optical components calls
for the development of several key technologies common to single
photonics.
\end{abstract}

\pacs{42.50.Ar,89.80.th,42.79.Ta,03.65.Bz}

\narrowtext

\section{Introduction}

The field of quantum computation has received tremendous new interest
since the recent result of Shor\cite{Shor94}, which shows the
possibility of using the non-local behavior of quantum mechanics to
factor integers in random polynomial time.  This is exponentially
better than is achievable on a comparable classical machine, with any
algorithm known today.

However, there is a catch.  Quantum computing (like quantum
cryptography) relies fundamentally on the processing of bits of
information which can be superpositions of logical one and zero.  As
long as the mutual coherence among a set of quantum bits
(qubits)\cite{Schum94} is preserved, they can simultaneously take on
more than one value, giving rise to a useful effect known as quantum
parallelism.  With sufficient cleverness, algorithms can be devised
which take advantage of this effect to solve some problems faster than
is possible with a classical computer.

The catch is that these qubits are ``Schr\"odinger cat'' states, which
are normally highly susceptible to collapse.  Whenever a qubit is
observed by an external agent (such as the environment\cite{Zurek93}),
coherence with other qubits in the system is partially lost due to the
collapse of its wavefunction.  This loss of coherence is accompanied
by a loss of information\cite{Land89} which is likely to cause a
malfunction of the quantum computer.  Thus, simply put, the
practicality of using quantum parallelism is crucially dependent on
our ability to build a machine which is sufficiently perfect and
isolated from its environment so as to preserve quantum coherence
throughout a calculation\cite{Unruh94}.

The key question upon which the feasibility of quantum computing
hinges is how difficult it is to maintain quantum coherence in a real
implementation.  This is very much a {\em system\/} issue, because to
succeed, not only must the logic devices be perfect, but also, the
scheme for their interconnection, and the method for preparing and
extracting the inputs and outputs of the computer.  Although
implementations of several quantum-mechanical logic
gates\cite{qmfg,Ober87} and general architectures\cite{Lloyd93,Div94}
have been proposed, no designs for a specific machine have yet
appeared in the literature, and therefore, it is unclear what the
minimum requirement is for realizing a complete system.  As a result,
it is also difficult to pin down what noise issues limit the
feasibility of maintaining quantum coherence in a complete quantum
computer.

The purpose of this article is to remedy this problem by proposing a
specific realization of a quantum computer which solves Deutsch's
problem\cite{deutsch92}.  Although the machine which we envision has
little practical use, it is a simple system which (1) demonstrates the
concept of quantum parallelism, and (2) delineates the desired quantum
behavior from the merely classical by using simple error correction.
The approach which we outline also describes several techniques which
we believe will be useful in constructing a more general purpose
machine.

We note in relation to the literature that many issues which arise in
the course of our discussion remain open questions.  In particular, we
do not attempt to address the problem of synthesizing a {\em
universal} quantum computer from some minimal set of logic
gates\cite{Lloyd95,Div94,Deutsch89}.  Neither are we particularly
interested in solving the full problem of quantum
error-correction\cite{Peres85,Zurek84}.  Instead, our concern is the
{\em reality} of quantum computing.  By focusing on the complete
design of a specific machine, we learn about realizability, operation,
and robustness -- system issues which are of principle concern in
understanding the impact of decoherence.  Our design of a simple
quantum computer using error correction provides a concrete and new
framework for analyzing the role of decoherence in quantum computing.

We begin by summarizing Deutsch's problem.  We then compare the
classical and quantum solutions to a simplified version of the
problem, and discuss how the required components may be realized.
This leads us to a design for a machine which we present in Section~4,
which is followed by an analysis of its error correcting ability in
Section~5.  We conclude with a discussion of the experimental
possibilities.

\section{Deutsch's Problem}

Deutsch's problem may be described as the following game.  Alice, in
Amsterdam, selects a number $x$ from $0$ to $2L-1$, and mails it in a
letter to Bob, in Boston.  Bob calculates some function $f(x)$ and
replies with the result, which is either $0$ or $1$.  Now, Bob has
agreed to use one of only two kinds of functions, either type (1),
which are constant for all values of $x$, or type (2), which are equal
to one for exactly half of all the possible $x$.  Alice's mission is
to determine with certainty which type of function Bob has chosen by
corresponding with him the fewest number of times.  How fast can she
succeed?

In the classical case, Alice may only send Bob one value of $x$ in
each letter.  At worst, Alice will need to query Bob at least $L+1$
times, since she may receive, e.g., $L$ zeros before finally getting a
one, telling her that Bob's function is type~2.  The best
deterministic classical algorithm she can use therefore requires $L+1$
queries.  Note that in each letter, Alice sends Bob $N$ bits of
information, where $N = \log_2 (2L)$.

Now add a new twist to the problem.  Suppose that Bob and Alice can
exchange quantum bits (instead of just classical bits), and
furthermore, Bob calculates $f(x)$ using a unitary transformation
$U_f$.  Alice can now get back more than one value of $f(x)$ from Bob
in a single query, while still exchanging only about $N$ bits.  For
example, Alice may send Bob an atom trap containing $N+1$ two-level
atoms.  The first $N$ atoms, representing $x$, are prepared in an
equal superposition of their excited and ground states, while the last
atom, a scratch-pad for the result $y=f(x)$, is put in its ground
state.  In Boston, Bob uses a sequence of electromagnetic pulses to
unitarily put atom $y$ in the state $f(x)$.  Note that $x$ is in a
superposition of all values $[0,2^N-1]$, and therefore, $y$ is left a
superposition of all possible values of $f(x)$.  However, when Alice
receives the reply, she can't achieve her mission simply by measuring
atom $y$, since that would collapse the superposition state and give
her only one result!

Instead, Alice must be more clever.  She gives $y$ a $\pi$ phase shift
relative to $x$, then sends the qubits once more to Bob.  This time,
Bob agrees to calculate ${U_f^\dagger}$ instead of $U_f$, i.e., he
inverts what he did before, leaving $y$ in its ground state.  Since $y$
and $x$ are entangled, this procedure also leaves the $N$ qubits of
$x$ with a special relative phase, such that those values of $x$ for
which $f(x)$ is even are be 180$^\circ$ out of phase with the others.
When Alice receives the result back from Bob, she can perform an
interference experiment to determine the type of Bob's function, with
certainty.  This is accomplished using only {\em two\/} queries.

The quantum algorithm followed by Alice in the latter case was devised
by Deutsch and Jozsa, and a more mathematical description can be found
in their article\cite{deutsch92}.  A schematic of the algorithm is
shown in Figure~\ref{fig:algorithm}.  This drawing, and our
description above highlight the two principle differences between
classical and quantum computing: (1) information is represented as
quantum bits, and (2) information interactions are performed using
unitary transformations.  These two changes allow Deutsch's problem to
be solved in ${\cal O}(N)$, rather than in ${\cal O}(\exp N)$ time.
In our example, physical distance was used to artificially elevate the
cost of calculating $f(x)$; this is not needed in general, where
$f(x)$ may be inherently difficult to calculate.  We shall study next
how qubits can be generated, manipulated, interacted, and measured.

\section{Components of a Quantum Computer}

The nature of the physical realization of the algorithm of
Figure~\ref{fig:algorithm} depends most on the representation chosen
for the quantum bit.  As we mentioned, two-level atoms are one
possibility.  Single electrons, solitons, magnetic flux quanta,
nuclear spins, and quantum dots are other possibilities which have
been considered.  We have chosen to represent qubits as single
photons, primarily because almost all the required components (for a
single photon quantum computer) exist today, but also because quantum
optics is a well-developed field in which noise is a thoroughly
understood subject.  However, we believe that there are some general
limitations governing all qubit representations, and our goal is to
try to elucidate those, so despite our use of quantum optics
terminology, it should be kept in mind that many of our conclusions
are applicable to other systems as well.

Given that we are using $\mbox{$|0\rangle$}$ (the vacuum state) and
$\mbox{$|1\rangle$}$ (the single photon state) to represent logical
zero and one, respectively, we must answer the following three
questions to construct our quantum computer to solve Deutsch's
problem:
\begin{itemize}
\item[(1)] 	How is a superposition state prepared?
\item[(2)]	What unitary transform is used to calculate $f(x)$?
\item[(3)]	What interference experiment is performed to
		    determine the final result?
\end{itemize}
That is, we need devices to perform the unitary operations $M$, $U_f$,
and $S$, and an architecture which provides a definite phase reference
so as to allow the final interference experiment to be performed.  We
now show how the traditional tools of optics can be used to fulfill
our needs.  We shall use beamsplitters, mirrors, phase shifters, and
Kerr media.

The first task is to create a superposition state.  It is possible in
principle to create the state
\begin{equation}
	\mbox{$|\psi\rangle$} = \frac{1}{\sqrt{2}}
	\left[\rule{0pt}{2.4ex}{ \mbox{$|0\rangle$} + \mbox{$|1\rangle$}
	}\right]
\end{equation}
but we have a simpler alternative.  The ordinary 50/50 optical
beamsplitter\cite{Campos89,Prasad87} acting on modes $a$ and $b$ is
described by the quantum operator $B$, shown in
Figure~\ref{fig:components}.  Let us label states as $\mbox{$|\sf a
b\rangle$}$.  A beamsplitter with input $\mbox{$|01\rangle$}$ gives
the output
\begin{equation}
	B \mbox{$|01\rangle$} = \frac{1}{\sqrt{2}}
	\left[\rule{0pt}{2.4ex}{ \mbox{$|01\rangle$} + \mbox{$|10\rangle$}
	}\right]
\,.
\end{equation}
Now comes our first trick.  Let us represent a single qubit by a pair
of modes, such that $\mbox{$|01\rangle$}$ and $\mbox{$|10\rangle$}$
are logical zero and one, respectively.  This {\em dual-rail\/}
representation of a logical state embeds an elementary form of error
correction which will be useful later.  With this representation, we
see that a simple beamsplitter can be used to generate the desired
superposition state of logical zero and one.

Next, we must calculate $f(x)$ using a unitary transform.  Since
$f(x)$ is a mapping from ${\cal Z}\rightarrow{\cal Z}_2$, we may
consider it to be calculable by an acyclic boolean circuit.  It is
therefore possible to implement it using a cascade of reversible logic
gates, such as the Fredkin gate\cite{Fred82}.  For example, consider
the two-bit Deutsch problem.  Here, $0\leq x < 4$, and there exist
eight possible functions which Bob may choose
(Table~\ref{tab:twobit}).  Two circuits which can be used to implement
$f(x)$ are shown in Figure~\ref{fig:circuits}B.  Also shown are
circuits for the one-bit problem, where $0\leq x<2$
(Table~\ref{tab:onebit}).  The reversible logic circuits correspond
directly to unitary operators which may be implemented as
quantum-mechanical transforms.  This is done simply by using a quantum
Fredkin gate in place of the classical one.

Note that this technique, of utilizing a reversible logic
implementation to determine the unitary operator necessary to
implement a classical function, is valid in the general case.  For
example, Shor's algorithm requires the calculation of $x^a\ {\rm mod}\
N$, for which the proper unitary transform may be arrived at through
analysis of the required reversible logic circuit.  Also note that we
have chosen the Fredkin gate in favor of the Toffoli gate, because
{\em conservative} invertible logic gates conserve the number of
``ones'' and therefore are possibly more amicable to qubit
representations where a logical one implies existence of some energy
packet (as will be the case for our system)\cite{reservoir}.

An optical realization of the quantum Fredkin gate
(Figure~\ref{fig:components}) has been proposed\cite{qmfg}, and is
understood well.  It is simply a nonlinear Mach-Zehnder
interferometer, with an external control signal which causes the
exchange of {\sf a} and {\sf b} by inducing a relative $\pi$ phase
shift in one arm via cross-phase modulation in the Kerr medium.  This
device may be viewed as a ``controlled beamsplitter,'' where the {\sf
c}-input determines the angle of a beamsplitter with inputs {\sf a}
and {\sf b}.  We shall let $\chi=\pi$, such that when
${{c^\dagger}c}=1$, the Fredkin operator $F$ acts on {\sf a} and {\sf
b} just like a beamsplitter with angle $\pi/2$, i.e.,
$F\mbox{$|101\rangle$} = -\mbox{$|011\rangle$}$ and
$F\mbox{$|011\rangle$} = \mbox{$|101\rangle$}$, where the state is
$\mbox{$|\sf abc\rangle$}$.  Note that when ${{c^\dagger}c}=0$, the
Fredkin operator is the identity, $F=I$.

Note that each of the components of our quantum computer, which
operate on {\em dual-rail\/} qubit representations, have a
corresponding description in the traditional picture of {\em
single-rail\/} qubit functions.  A two-input beamsplitter operating on
modes {\sf a} and $\overline{\sf a}$ is equivalent to Deutsch's
one-input $\sqrt{\mbox{\sc not}}$ gate\cite{Deutsch85} acting on the
qubit represented by the pair $\{{\sf a},\overline{\sf a}\}$.
Similarly, three three-input Fredkin gates acting on modes {\sf a},
$\overline{\sf a}$, {\sf b}, $\overline{\sf b}$, {\sf c}, and
$\overline{\sf c}$ can perform any thee-input Toffoli gate transform
on the three quibits represented by the pairs $\{{\sf a},\overline{\sf
a}\}$, $\{{\sf b},\overline{\sf b}\}$, and $\{{\sf c},\overline{\sf
c}\}$\cite{Chuang94qmlg}; in this sense, the Fredkin gate is close to
DiVincenzo's ``controlled-rotation'' gate\cite{Div94}.  Incidentally,
since it has been shown that these traditional gates are
``universal,'' in the sense that they can be cascaded to synthesize
any arbitrary quantum computing device, it follows that our component
set is also universal.

One more unitary operator which is needed is the phase shift $S$
performed by Alice after receiving the first letter back from Bob.
This is accomplished using a $\pi$ phase delay.  Finally, the task of
interference and measurement can be performed by using an
interferometer and ideal photon counters.  Alice can create and
decorrelate superpositions using beamsplitters and communicate to Bob
by sending him photons; and Bob can calculate his function using
Fredkin gates.  Thus, the Deutsch-Jozsa quantum algorithm may be
implemented using the traditional components of quantum optics.  This
viewpoint will be useful in analyzing the physics of our machine as we
assemble it in the following section.

\section{The Machine}

The one-bit Deutsch problem is the simple case where Alice sends Bob a
value of $x=0$ or $x=1$, and Bob replies with $f(x)$, where he has
chosen one of the four functions shown in Table~\ref{tab:onebit}.
Clearly, in the classical case, Alice can achieve her goal of
determining the type of Bob's function by sending Bob just two
queries.  The quantum solution can be achieved with the same number of
queries, so there is no time advantage in this case.  However, it is
worthwhile to consider precisely how the quantum algorithm is
implemented to understand the role which quantum coherence plays.

The machine which we propose is diagrammed in
Figure~\ref{fig:machine}.  The general operation is as follows.  Alice
prepares two qubits $\{{\sf a},{\sf b}\}$ and $\{{\sf c},{\sf d}\}$,
each of which is represented by a dual-rail single-photon eigenstate.
Operationally, this means that she sends single photon eigenstates
simultaneously into modes {\sf d} and {\sf b}, and the vacuum state
into the other two.  The $\{{\sf c},{\sf d}\}$ qubit is passed through
a $\sqrt{\mbox{\sc not}}$ gate which implemented by a beamsplitter to
prepare a value of $x$ which is in a 50/50 superposition of $0$ and
$1$.  This qubit is passed along with the scratch-pad qubit $\{{\sf
a},{\sf b}\}$ to Bob.  Bob uses a quantum Fredkin gate and three
classical switches to perform his calculation, and returns $f(x)$ in
the scratch-pad.  Alice gives the result a relative $\pi$ phase shift,
then allows Bob to invert his first transform.  Finally, Alice sends
the $\{{\sf c},{\sf d}\}$ qubit through a final beamsplitter, and
measures the number of photons she receives in all four modes.  In the
absence of error, the detector for mode {\sf d} tells Alice the type
of Bob's function with certainty, from a single execution of the
machine.

Let us now analyze the behavior of this machine by calculating the
states $\mbox{$|\psi_i\rangle$}$, defined as
\begin{eqnarray}
	\mbox{$|\psi_0\rangle$} &=& \mbox{Alice's initial state}
\nonumber\\
	\mbox{$|\psi_1\rangle$} &=& \mbox{Superposition state sent to Bob}
\nonumber\\
	\mbox{$|\psi_2\rangle$} &=& \mbox{Result returned to Alice the
		first time}
\nonumber\\
	\mbox{$|\psi_3\rangle$} &=& \mbox{Phase shifted state sent back to Bob}
\nonumber\\
	\mbox{$|\psi_4\rangle$} &=& \mbox{Result returned to Alice the
	second time}
\nonumber\\
	\mbox{$|\psi_5\rangle$} &=& \mbox{Alice's final state, after
	decorrelation}
\nonumber\,.
\end{eqnarray}
We shall label the states as $\mbox{$|\sf abcd\rangle$}$, and use the
fact that $S$ acts on mode {\sf a}, $B$ acts on {\sf c} and {\sf d},
and $U_f$ acts on {\sf a}, {\sf b}, and {\sf c}.  We may think of mode
{\sf c} of state $\mbox{$|\psi_1\rangle$}$ as the value of $x$
prepared by Alice to send to Bob, and mode {\sf a} of state
$\mbox{$|\psi_2\rangle$}$ as the value of $f(x)$ returned by Bob.
When $k_1k_0=00$, the {\sf c} and {\sf d} modes are completely
decoupled from the lower circuit.  Using our beamsplitter convention,
the states are thus
\begin{eqnarray}
	\mbox{$|\psi_0\rangle$} &=& \mbox{$|0101\rangle$}
\\
	\mbox{$|\psi_1\rangle$} &=& B\mbox{$|\psi_0\rangle$}
	=
	\frac{1}{\sqrt{2}}\left[\rule{0pt}{2.4ex}{\mbox{$|0101\rangle$}
	+\mbox{$|0110\rangle$}}\right]
\\
	\mbox{$|\psi_4\rangle$} &=& \mbox{$|\psi_3\rangle$} =
	\mbox{$|\psi_2\rangle$} = \mbox{$|\psi_1\rangle$}
\\
	\mbox{$|\psi_5\rangle$} &=& {B^\dagger}\mbox{$|\psi_4\rangle$}
	= \mbox{$|0101\rangle$}
\,.
\end{eqnarray}
This is the expected result, because {\sf c} and {\sf d} form an
independent, balanced Mach-Zehnder interferometer, and since the
control input to the Fredkin gate is zero, no switching occurs, and
the output state is the same as the input.  Note that the result is a
pure state, and so the photon number measurement result is not
stochastic.  If the function chosen by Bob is $k_1k_0=01$, the result
is similar; this time, the phase shift $S$ interacts with the photon
input to mode {\sf b}, giving us
\begin{eqnarray}
	\mbox{$|\psi_0\rangle$} &=& \mbox{$|0101\rangle$}
\\
	\mbox{$|\psi_1\rangle$} &=&
	\frac{1}{\sqrt{2}}\left[\rule{0pt}{2.4ex}{\mbox{$|0101\rangle$}
	+\mbox{$|0110\rangle$}}\right]
\\
	\mbox{$|\psi_2\rangle$} &=&
	\frac{1}{\sqrt{2}}\left[\rule{0pt}{2.4ex}{\mbox{$|1001\rangle$}
	+\mbox{$|1010\rangle$}}\right]
\\
	\mbox{$|\psi_3\rangle$} &=& S \mbox{$|\psi_2\rangle$}
	=
	\frac{1}{\sqrt{2}}\left[\rule{0pt}{2.4ex}{-\mbox{$|1001\rangle$} -
	\mbox{$|1010\rangle$}}\right]
\\
	\mbox{$|\psi_4\rangle$} &=&
	\frac{1}{\sqrt{2}}\left[\rule{0pt}{2.4ex}{-\mbox{$|0101\rangle$} -
	\mbox{$|0110\rangle$}}\right]
\\
	\mbox{$|\psi_5\rangle$} &=& {B^\dagger}\mbox{$|\psi_4\rangle$}
	= -\mbox{$|0101\rangle$}
\,.
\end{eqnarray}
Both these results are trivial, since whenever $k_1=0$, the result
returned by Bob, $f(x)$, is {\em independent\/} of $x$.

However, a nontrivial output results when $k_1=1$.  Consider
$k_1k_0=10$.  Here, Bob's transform $U_{f_{10}}=F$ is a Fredkin gate
acting on ${\sf a}$, ${\sf b}$, and ${\sf c}$, and we get
\begin{eqnarray}
	\mbox{$|\psi_0\rangle$} &=& \mbox{$|0101\rangle$}
\\
	\mbox{$|\psi_1\rangle$} &=& B\mbox{$|\psi_0\rangle$}
	=
	\frac{1}{\sqrt{2}}\left[\rule{0pt}{2.4ex}{\mbox{$|0101\rangle$}
	+\mbox{$|0110\rangle$}}\right]
\\
	\mbox{$|\psi_2\rangle$} &=& U_{f_{10}}\mbox{$|\psi_1\rangle$}
	= \frac{1}{\sqrt{2}}
	\left[\rule{0pt}{2.4ex}{\mbox{$|0101\rangle$}
	+\mbox{$|1010\rangle$}}\right]
\\
	\mbox{$|\psi_3\rangle$} &=& S\mbox{$|\psi_2\rangle$}
	=
	\frac{1}{\sqrt{2}}\left[\rule{0pt}{2.4ex}{\mbox{$|0101\rangle$} -
	\mbox{$|1010\rangle$}}\right]
\\
	\mbox{$|\psi_4\rangle$} &=& {U^\dagger}_{f_{10}}
	\mbox{$|\psi_3\rangle$}
	= \frac{1}{\sqrt{2}}
	\left[\rule{0pt}{2.4ex}{\mbox{$|0101\rangle$} -
	\mbox{$|0110\rangle$}}\right]
\\
	\mbox{$|\psi_5\rangle$} &=& {B^\dagger}\mbox{$|\psi_4\rangle$}
	= - \mbox{$|0110\rangle$}
\label{eq:kozresult}
\,.
\end{eqnarray}
This result can be understood by realizing that if the control signal
input to a quantum Fredkin gate is a superposition state, then the
outputs will also be superposition states.  Thus, the state
$\mbox{$|\psi_2\rangle$}$ returned by Bob leaves $y$ in a
superposition state, and since the phase shift $S$ has an effect only
when its input is $\mbox{$|1\rangle$}$ (i.e., not the vacuum), it
``filters'' out and marks those cases where $f(x)$ has odd parity.
This nontrivial result is obtained by virtue of the quantum coherence
between all four states.  The result for $k_1k_0=11$ is similar:
\begin{eqnarray}
	\mbox{$|\psi_0\rangle$} &=& \mbox{$|0101\rangle$}
\\
	\mbox{$|\psi_1\rangle$} &=& B\mbox{$|\psi_0\rangle$}
	=
	\frac{1}{\sqrt{2}}\left[\rule{0pt}{2.4ex}{\mbox{$|0101\rangle$}
	+\mbox{$|0110\rangle$}}\right]
\\
	\mbox{$|\psi_2\rangle$} &=& U_{f_{11}}\mbox{$|\psi_1\rangle$}
	= \frac{1}{\sqrt{2}}
	\left[\rule{0pt}{2.4ex}{\mbox{$|1001\rangle$}
	+\mbox{$|0110\rangle$}}\right]
\\
	\mbox{$|\psi_3\rangle$} &=& S\mbox{$|\psi_2\rangle$}
	=
	\frac{1}{\sqrt{2}}\left[\rule{0pt}{2.4ex}{-\mbox{$|1001\rangle$} +
	\mbox{$|0110\rangle$}}\right]
\\
	\mbox{$|\psi_4\rangle$} &=& {U^\dagger}_{f_{11}}
		\mbox{$|\psi_3\rangle$}
	= \frac{1}{\sqrt{2}} \left[\rule{0pt}{2.4ex}{\mbox{$|0110\rangle$} -
		\mbox{$|0101\rangle$}}\right]
\\
	\mbox{$|\psi_5\rangle$} &=& {B^\dagger}\mbox{$|\psi_4\rangle$}
	= \mbox{$|0110\rangle$}
\,.
\end{eqnarray}
Note that the output is very different when $k_1$ is zero or one.  Let
$z$ be the measurement result for mode {\sf d}.  When $k_1=0$, the
result is $z=1$, and Alice's correct conclusion is that Bob's function
is type~1.  Likewise, when $k_1=1$, Alice finds that $z=0$, and
concludes that Bob's function is type~2.

Another way to understand physically what is happening is to reduce
the circuit by breaking the abstraction barrier around Bob's
apparatus, and taking advantage of the fact that a $\pi$ phase shift
sandwiched between two beamsplitters is just a crossover switch.  We
consider the $k_1k_0=10$ case, where the circuit reduces to become
that shown in Figure~\ref{fig:reduced}A.  We have two interferometers
linked by Kerr media; in the bottom interferometer, the photon is
split at the first beamsplitter.  If it takes the upper path, then it
causes a $\pi$ phase shift in mode {\sf c} via cross-phase modulation
in the first Kerr medium.  Alternatively, if the photon takes the
bottom path, it also causes a $\pi$ phase shift in {\sf c}, this time
through the second Kerr medium.  Either way, the result is the same;
the upper interferometer is unbalanced by $\pi$, and thus its inputs
are exchanged to give the outputs.  This explains why the output is
$\mbox{$|\psi_5\rangle$}=\mbox{$|0110\rangle$}$ in
Eq.(\ref{eq:kozresult}).  Note the usefulness of the Everett
many-worlds interpretation of quantum mechanics\cite{Everett57} in
explaining the operation of this quantum computer.  Another
interesting observation is that if $k_1=1$, then inserting and
removing the phase shift $S$ should have the effect of turing $k_1$ on
and off.  This effect is the signature of quantum parallelism in our
apparatus.

%

Finally, it is interesting to consider what happens if classical
operation of this machine is attempted.  If a coherent state
$\mbox{$|\alpha\rangle$}$ is used to represent logical one, and the
vacuum $\mbox{$|0\rangle$}$ as logical zero, the machine will fail in
the following way: the measurement results will be independent of
whether $S$ is in-place or removed.  Consider the $k_1k_0=10$ case,
and simplify the circuit to the two circumstances shown in
Figure~\ref{fig:reduced}.  Now, it is well known that the outputs of a
beamsplitter fed with a coherent state and a vacuum input are coherent
states with half the expected photon number,
\begin{equation}
	B\mbox{$|0,\alpha\rangle$} =
		\mbox{$|\alpha/\sqrt{2},-\alpha/\sqrt{2}\rangle$}
\,,
\end{equation}
since this is just the expected classical behavior.  In this case,
both arms of the lower interferometer will contain the same number of
photons, so the photons in mode {\sf c} will receive the same
cross-phase modulation in both cases.  When $S$ is in-place, {\sf c}
will get a phase shift once from {\sf b} and once from {\sf a}, and
when $S$ is removed, {\sf c} will be phase shifted twice by {\sf b}.
Since the amount of shift is the same in either case, the measurement
result is independent of presence of $S$.

This shows that quantum parallelism does not occur in our machine
under classical operation.  This is not a surprising result, since a
beamsplitter does not create a Schr\"odinger cat state of
$\mbox{$|0\rangle$}$ and $\mbox{$|\alpha\rangle$}$ from a coherent
state input.

\section{Error Correction}

An important feature of our simple quantum computer is its use of a
dual-rail qubit representation.  Given correct input preparation, we
expect at all times that a single photon exists in {\em either\/} mode
{\sf c} or {\sf d}, but not both; likewise for modes {\sf a} and ${\sf
b}$.  This feature allows us to detect certain cases when information
is lost from the computer, and reject the faulty data.  Although this
error correction scheme is simple-minded and does not solve the
general quantum error correction problem, it is simple to implement,
and effective in reducing the probability of error, as we shall see in
this section.

Because the machine operates deterministically under perfect
conditions, error correction is easy.  If the measurement result for
the four modes ever changes without any change of the inputs or the
switch conditions, then somewhere, a random process must be
interacting with the qubits in the machine.  For example, measurement
of a total of zero or one photons at the output is indicative of a
loss process, while measurement of more than two photons suggests some
error in preparation of the inputs.  Assuming that input preparation
is always perfect, we may correct for random errors by rejecting all
executions which result in one of $|0000\rangle$, $|0001\rangle$,
$|0010\rangle$, $|0100\rangle$, or $|1000\rangle$.  We may also reject
$|1010\rangle$ and $|1001\rangle$, since we know {\em a priori\/} that
the scratch-pad (qubit $\{{\sf a},{\sf b}\}$) should remain logically
unchanged.  When rejection occurs, we perform a re-trial execution.

Let us now consider a specific decoherence model.  The Kerr medium
used by Bob in his quantum Fredkin gate is experimentally known to be
lossy\cite{Watanabe90}, and we may model this by inserting a loss
mechanism in modes {\sf b} and {\sf c}.  Without loss of generality,
we consider just the $k_1k_0=10$ case, and imagine having loss occur
only during the second instantiation of Bob's apparatus.
Specifically, just as before, we have
\begin{equation}
	|\psi_3\rangle = SU_{f_{10}}B |0101\rangle
\end{equation}
as the state sent by Alice to Bob in her second communication.  We now
dismantle Bob's apparatus; in the absence of decoherence, Bob performs
the transform $U_{f_{10}}= B_{\rm ab} K_{\rm bc}
B^\dagger_{\rm ab}$, where $B_{\rm ab}$ is the usual 50/50
beamsplitter acting on modes {\sf a} and {\sf b}, and $K_{\rm bc}
=
\exp[i \pi b^\dagger b c^\dagger c]$ is the Kerr operator acting on
modes {\sf b} and {\sf c}.  However, we shall consider instead
$\tilde{U}_{f_{10}}= B_{\rm ab} \Gamma_{\rm b}\Gamma_{\rm
c} K_{\rm bc} B^\dagger_{\rm ab}$, where $\Gamma_i$ is a {\em
non-unitary\/} amplitude damping operator acting on mode $i$.  The
formal operation of $\Gamma_i$ is best described by its action on a
general single qubit density matrix,
\begin{equation}
	\Gamma_i
	   \left[
	    \begin{array}{cc}
		\rho_{00}  & \rho_{01} \\
		\rho_{10}  & \rho_{11} \\
	    \end{array}
	    \right]
	\Gamma_i^\dagger
 =
	   \left[
	    \begin{array}{cc}
		  \rho_{00}  + (1-e^{-\gamma}) \rho_{11}
		& e^{-\gamma/2} \rho_{01} \\
		  e^{-\gamma/2} \rho_{10}
		& e^{-\gamma}   \rho_{11} \\
	    \end{array}
	    \right]
\,.
\label{eq:damping}
\end{equation}
In other words, $\Gamma_i$ describes the amplitude damping due to a
Caldeira-Leggett type coupling\cite{Caldeira83} of mode $i$ to the
environment, with coupling constant $\gamma$.  We concern ourselves
only with the reduced density matrix of the system here; a good
description of this procedure can be found in standard quantum-optics
textbooks\cite{Louisell}.

The calculation of the output result is straightforward using density
matrices.  We get
\begin{eqnarray}
	|\psi_{3a}\rangle &=& B_{\rm ab} |\psi_3\rangle
\\
	\rho_{3a} &=& |\psi_{3a}\rangle \langle \psi_{3a}|
\\
	\rho_{3b} &=& \Gamma_{\rm b}\Gamma_{\rm c}
				\rho_{3a} \Gamma_{\rm c}^\dagger
					\Gamma_{\rm b}^\dagger
\\
	\rho_{3d} &=& B_{\rm ab} K \rho_{3d} K^\dagger
			B_{\rm ab}^\dagger
\\
	\rho_4 &=& B^\dagger \rho_{3d} B
\,,
\end{eqnarray}
where the density matrix $\rho_{3a}$ describes the input to the loss
medium, $\rho_{3b}$ is the input to the Kerr medium (calculated using
Eq.~\ref{eq:damping}), $\rho_{3d}$ is the output of Bob's apparatus,
and $\rho_4$ is the final output.  The diagonal elements of $\rho_4$
give us the final measurement result probabilities.  Physically, we
expect errors to occur because the loss of photons results in the
possibility of the second Fredkin gate failing to switch.  Thus, loss
either causes an incorrect total output photon count, or results in
the incorrect location of an output photon.

Without error correction, we simply look at the measurement result for
mode {\sf d}.  Since the expected result is that $z=0$ for the
$k_1k_0=10$ case, we find that the error probability is
\begin{equation}
	P_{\mbox{\sc noec}} =
	\frac{1}{4}
	\left[\rule{0pt}{2.4ex}{
					1 + e^{-\gamma} - 2e^{-3\gamma/2}
	}\right]
\,.
\end{equation}
On the other hand, if we perform error correction by rejecting all
illegal results, then the error probability is given by the relative
probability of getting $|0101\rangle$ (the wrong answer) to
$|0110\rangle$ (the right answer),
\begin{equation}
	P_{\mbox{\sc ec}} =
	\frac{1}{2}
	\left[\rule{0pt}{2.4ex}{
					1 - {\rm sech}\, \frac{\gamma}{2}
	}\right]
\,.
\end{equation}
The dramatic improvement in our error rate given by use of the
dual-rail qubit error correction scheme is shown in
Figure~\ref{fig:biterror}.  Work is currently in progress to extend
these results to consider other noise sources, such as phase
randomization.

It is possible to generalize our results to the $N$-bit Deutsch
problem, using the techniques outlined in the previous two sections,
although we shall not do so here.  Rather, let us summarize the
findings from the study of our simple quantum computer: (1) the
concept of quantum parallelism, demonstrated through the simultaneous
calculation of $f(x)$ for two values of $x$, is not in conflict with
any fundamental principle of physics, or any fundamental source of
noise that is apparent in our system, and (2) rudimentary error
correction using a dual-rail qubit representation is simple to apply
to a quantum computer, and indeed can be effective in indicating
coherence loss or improper input preparation.  These advances are
hopeful signs of the eventual practicality of quantum computing.

\section{Conclusion}

The experimental realization of a quantum computer is a difficult
proposition.  By definition, unitary evolution requires complete
isolation from the environment.  However, at the same time, it must
be possible for qubits to interact with each other, so that
information processing can occur.  This dilemma goes to the heart of a
tradeoff that is central to the practicality of quantum computing.


We chose to use single photons as representations of a qubit, in part
because it is easy to create superpositions of single photons using a
normal beamsplitter.  However, it turns out that it is difficult to
find a nonlinear optical material with a $\chi^{(3)}$ coefficient
sufficiently strong to allow two single photons to give each other
$\pi$ cross phase modulation.  In contrast, it is easy to cross-phase
modulate two single electrons, via the Coulomb
interaction\cite{Kitagawa91}, but difficult to fabricate a 50/50
electron beamsplitter shorter than the dephasing length in a
high-mobility semiconductor electron gas.  The tradeoff is the
interaction strength; it seems that in general, if bits strongly
interact, then it is easy to make them process information, but
difficult to put them into superposition states.

Another general observation comes from contemplating the structure of
our quantum computer.  There are three interferometers in this simple
one-bit machine!  The problem is that quantum computing involves the
storage and manipulation of information in canonically conjugate
degrees of freedom.  For example, in our apparatus, information is
encoded both in the photon number (in each mode) and the phase of the
photon.  Interferometers are used to convert between the two
representations.  This is fine, in our system, because it is feasible
to construct stable optical interferometers.  However, if an
alternate, massive representation of a qubit were chosen, then it
would rapidly become difficult to build stable interferometers,
because of the shortness of typical de Broglie wavelengths.

Both of the above problems deal with coherence.  There is also the
issue of timing.  The quantum computer envisioned here is ballistic.
Although the machine we present is, in principle, perfectly
reversible, we have implicitly assumed that no scattering takes place
within the system, because such effects would lead to timing jitter
which would cause the malfunctioning of the machine.  That is because
the logical state of our machine is distributed among four modes, and
we cannot deal with effects which cause temporal synchronization to be
lost.  The only solution we have is that given to us by our simple
error correction method; in the event of a detected error, throw out
the execution trial and try again.

Despite these problems, we believe that Nature favors quantum
computing with single photon states in several ways.  First, it is
very easy to create superposition states using a beamsplitter.  These
states have been called Schr\"odinger kittens because of their
robustness compared to macroscopic superposition states which are more
massive.  Also, transformations such as the phase shift $S$ have
simple realizations, because ${a^{\dagger}} a$ is the number operator
for a single photon, rather than for something macroscopic.  These
features suggest that single photons (or single electrons) are
appropriate physical realizations of quantum bits.

Furthermore, we believe that imminent technological advances in the
area of single photonics may provide some impetus to the realization
of our machine.  In particular, we suggest that the single photon
turnstile device\cite{Imamoglu94} may be the solution for generating a
quantum bit source with high spectral purity and a well defined clock.
This would give us delocalized states with a high Kerr interaction
cross-section, and robustness against timing errors.  Also, we hope
for a new generation of single-photon detectors, such as the
single-photon gate FET\cite{Imamoglu93} and new avalanche
photodetectors\cite{Kwiat94}.  Finally, we look forward to new
nonlinear optical interactions which may give us single-photon driven
switches by coherently converting a photon to and from some other
particle (e.g., the exiton-polariton) which has a larger nonlinear
interaction strength\cite{Jacobson94}.

Realization of our simple quantum computer using optical components is
attractive because of the simplicity of our proposal.  Because of
mirror symmetry, only one quantum logic gate need be implemented.
Furthermore, as a practical initial test of quanutum parallelism (and
the feasibility of maintaining quantum coherence through a nonlinear
medium), Kerr media with $\chi<\pi$ may be used.  In this case,
insertion and removal of the phase shift $S$ will still give a
statistical signature showing whether classical or quantum operation
has been achieved.

Our design of a simple quantum computer has laid a foundation upon
which more complicated and general purpose systems may be formulated.
By describing quantum computation in terms of the traditional tools of
quantum optics, and by introducing a system complete with rudimentary
error correction, we have constructed an simple framework for
analyzing the impact of decoherence, and evaluating the reality of
quantum computation.  We hope that our work will lead to a future
experiment to demonstrate the practicality of quantum computing.

\section{Acknowledgements}

We would like to thank B. Yurke for suggesting the use of optical
beamsplitters to create superposition states.  Thanks also to
J. Jacobson and W. Zurek for helpful discussions, and to the referee
for constructive comments.  The work of ILC was supported by the John
and Fannie Hertz Foundation.




\begin{table}
\begin{tabular}{cccccccccc}
$x_1$ & $x_0$ & $f_{000}$ & $f_{001}$ & $f_{010}$ & $f_{011}$
	& $f_{100}$ & $f_{101}$ & $f_{110}$ & $f_{111}$ \\ \hline
0 & 0	& 0 & 1 & 0 & 1 & 0 & 1 & 0 & 1 \\
0 & 1	& 0 & 1 & 1 & 0 & 0 & 1 & 1 & 0 \\
1 & 0	& 0 & 1 & 0 & 1 & 1 & 0 & 1 & 0 \\
1 & 1	& 0 & 1 & 1 & 0 & 1 & 0 & 0 & 1 \\
\end{tabular}
\caption{All possible functions $f_{k_2k_1k_0}(x)$ for $0\leq x < 4$.
$f_{000}$ and $f_{001}$ are type 1, while the rest are type 2.}
\label{tab:twobit}
\end{table}

\begin{table}
\begin{tabular}{ccccc}
$x$	& $f_{00}$	& $f_{01}$	& $f_{10}$	& $f_{11}$ \\ \hline
 0  	&  0  		&  1		& 0		& 1  	\\
 1  	&  0  		&  1  		& 1		& 0	\\
\end{tabular}
\caption{All possible functions $f_{k_1k_0}(x)$ for $0\leq x < 2$.
$f_{00}$ and $f_{01}$ are type 1, while the rest are type 2.}
\label{tab:onebit}
\end{table}



\begin{figure}[p]
\caption{Algorithm for solving Deutsch's problem using a quantum computer.}
\label{fig:algorithm}
\end{figure}

\begin{figure}[p]
\caption{Unitary transforms for the components of our quantum
	computer.  The operators $a$ and ${a^{\dagger}}$ are the usual
	annihilation and creation operators.}
\label{fig:components}
\end{figure}

\begin{figure}[p]
\caption{Boolean logic (left) and reversible logic (right) circuits for the
	calculation of the (A) one-bit and (B) two-bit functions
	$f(x)$.  $k_0$, $k_1$, and $k_2$ control the classical
	switches which determine the function calculated.  They are
	set (secretly) by Bob.}
\label{fig:circuits}
\end{figure}

\begin{figure}[p]
\caption{Complete quantum computer system used to solve the one-bit
	 Deutsch problem.  The apparatus in the dashed box is used by
	Bob to calculate $f_k(x)$, and everything else belongs to
	Alice.  In principle, it is not necessary to send mode {\sf d}
	to Bob, although it may simplify the implementation in practice.}
\label{fig:machine}
\end{figure}

\begin{figure}[p]
\caption{Simplified versions of the quantum computer circuit when
	$k_1k_0=10$, Bob's apparatus is merged in, and (A) the $\pi$~phase
	shift $S$ is in-place, or (B) $S$ is removed.}
\label{fig:reduced}
\end{figure}

\begin{figure}[p]
\caption{Error probability for the final measurement result in the
	$k_1k_0=10$ case, with and without error correction (lower and upper
	curves).  As loss increases to infinity, the error correction scheme
	becomes ineffective because the photons become localized in an arm of
	the interferometer, but for small $\gamma$, the improvement is
	substantial; $P_{\mbox{\sc noec}}\sim
	\gamma/2$ and $P_{\mbox{\sc ec}}\sim \gamma^2/16$, where loss is $4.34
	\gamma$ [dB].}
\label{fig:biterror}
\end{figure}



\begin{thebibliography}{10}

\bibitem{Shor94}
P. Shor,  in {\em Proc. 35$^{th}$ Annual Symposium on Foundations of Computer
  Science} (IEEE Press, USA, 1994).

\bibitem{Schum94}
B. Schumacher, Workshop on Quantum Computing and Communication, Gaithersburg,
  MD, August 18-19  (1994).

\bibitem{Zurek93}
W.~H. Zurek, Progress of Theoretical Physics {\bf 89},  281  (1993).

\bibitem{Land89}
R. Landauer, J. Stat. Phys. {\bf 54},  516  (89).

\bibitem{qmfg}
Y. Yamamoto, M. Kitagawa, and K. Igeta,  in {\em Proc. 3rd Asia-Pacific Phys.
  Conf.} (World Scientific, Singapore, 1988); G.~J. Milburn, Phys.
Rev. Lett. {\bf 62},  2124  (1989); I. Chuang, unpublished  (1993).

\bibitem{Unruh94}
W.~G. Unruh, UBC preprint, hep-th/9406058  (1994).

\bibitem{Ober87}
K. Obermayer, G. Mahler, and H. Haken, Phys. Rev. Lett. {\bf 58},  1792
  (1987).

\bibitem{Lloyd93}
S. Lloyd, Science {\bf 261},  1569  (1993).

\bibitem{Div94}
D.~P. DiVincenzo, Workshop on Quantum Computing and Communication,
  Gaithersburg, MD, August 18-19  (1994).

\bibitem{deutsch92}
D. Deutsch and R. Jozsa, Proc. R. Soc. Lond. A {\bf 439},  553  (1992).

\bibitem{Lloyd95}
S. Lloyd, preprint, Information Sciences, Dept. of Mech. Eng., MIT  (1995).

\bibitem{Deutsch89}
D. Deutsch, Proc. R. Soc. Lond. A {\bf 425},  73  (1989).

\bibitem{Peres85}
A. Peres, Phys. Rev. A {\bf 32},  3266  (1985).

\bibitem{Zurek84}
W.~H. Zurek, Phys. Rev. Lett. {\bf 53},  391  (1984).

\bibitem{Campos89}
Campos, Saleh, and Tiech, Phys. Rev. A {\bf 40},  1371  (1989).

\bibitem{Prasad87}
Prasad, Scully, and Martienssen, Optics Comm. {\bf 62},  139  (1987).

\bibitem{Fred82}
E. Fredkin and T. Toffoli, Int. J. of Theor. Physics {\bf 21},  219  (1982).

\bibitem{reservoir}
Energy exchange {\em within} the system need not imply information loss;
  however, energy loss {\em to the environment} -- which may be thought of as
  an infinite ensemble harmonic oscillators with no memory -- necessarily
  implies information loss from the system. If the unitary logic gate operator
  fails to commute with the system Hamiltonian, then when the logic interaction
  is switched on and off energy must flow to and from an external reservior,
  which is by default the environment.

\bibitem{Deutsch85}
D. Deutsch, Proc. R. Soc. Lond. A {\bf 400},  97  (1985).

\bibitem{Chuang94qmlg}
I.~L. Chuang, J. Jacobson, , and Y. Yamamoto, Submitted to Phys. Rev. Let.
  (1994).

\bibitem{Everett57}
H.~Everett~III, Rev. Mod. Phys. {\bf 29},  454  (1957).

\bibitem{Watanabe90}
K. Watanabe and Y. Yamamoto, Phys. Rev. A {\bf 42},  1699  (1990).

\bibitem{Caldeira83}
A.~O. Caldeira and A.~J. Leggett, Ann. Phys. {\bf 149},  374  (1983).

\bibitem{Louisell} See, for example,
W.~H. Louisell, {\em Quantum Statistical Properties of Radiation} (Wiley, New
  York, 1973), chapter 6.

\bibitem{Kitagawa91}
M. Kitagawa and M. Ueda, Phys. Rev. Lett. {\bf 67},  1852  (1991).

\bibitem{Imamoglu94}
A. Imamoglu and Y. Yamamoto, Phys. Rev. Lett. {\bf 72},  210  (1994).

\bibitem{Imamoglu93}
A. Imamoglu and Y. Yamamoto, Int. J. Mod. Phys B {\bf 7},  2065  (1993).

\bibitem{Kwiat94}
P.~G. Kwiat {\it et~al.}, Applied Optics {\bf 33},  1844  (1994).

\bibitem{Jacobson94}
J. Jacobson {\it et~al.},  in {\em Extended Abstracts of the 1994 Int. Conf. on
  Solid State Devices and Materials, Yokohama} (The Japan Soc. of Appl. Phys.,
  Tokyo, 1994).

\end{thebibliography}
\end{document}